\title{STRONG-COUPLING EXPANSION \\ FOR THE HUBBARD MODEL \\ 
IN ARBITRARY DIMENSION\\ USING SLAVE BOSONS}
\author{ P.J.H. Denteneer\thanks{e-mail address: 
pjhdent@rulkol.leidenuniv.nl} \\
        Instituut--Lorentz, University of Leiden,\\
        P. O. Box 9506, 2300 RA Leiden, The Netherlands}
\date{}
\documentclass[12pt]{article}
\begin{document}
\hoffset=-0.0 true cm
\voffset=-2.0 true cm
\maketitle
\vspace{1cm}
\begin{abstract}
A strong-coupling expansion for the antiferromagnetic phase of the
Hubbard model is derived in the framework of the
slave-boson mean-field approximation. The expansion can be obtained
in terms of moments of the density of states of freely hopping
electrons on a lattice, which in turn are obtained for hypercubic
lattices in arbitrary dimension. The expansion is given for the case
of half-filling and for the energy up to fifth order in the ratio of 
hopping integral $t$ over on-site interaction $U$, but
can straightforwardly be generalized to the non-half-filled case 
and be extended to higher orders in $t/U$.
For the energy the expansion is found to have an accuracy of better
than $1 \%$ for $U/t \geq 8$. A comparison is given with an earlier
perturbation expansion based on the Linear Spin Wave approximation
and with a similar expansion based on the Hartree-Fock approximation.
The case of an infinite number of spatial dimensions is discussed.
\end{abstract}
\vspace{1.0cm}
Short title: Strong-coupling expansion for the Hubbard model \\[3.0cm]
INLO-PUB 23/95; January 1996; to be published in Physical Review B.\\
%e-mail: pjhdent@rulkol.leidenuniv.nl

\newpage
 
% Section 1: General intro
\section{Introduction}
If the physics of strongly correlated fermions is to be described by
the Hubbard model \cite{Hub}, electronic structure calculations for
the cuprates indicate that the relevant parameter regime is where
the on-site interaction $U$ is comparable to 
or larger than the bandwidth $W$ of freely
hopping electrons \cite{HybMarSaw}. A natural strategy to try to
describe
this regime is to approach it from the two limiting cases of
weak-coupling ($U \ll W$) and strong-coupling ($U \gg W$).
The simplest mean-field approximation for the Hubbard model, the
Hartree-Fock approximation (HFA), is at first sight a 
weak-coupling approximation
although, with some care, also a reasonable description of the
strong-coupling regime can be given \cite{SWZ}--\cite{PDEPL}.
Strong-coupling approaches have mainly been devoted to the 
one-dimensional case (see e.g. \cite{CarBaer}) and often take 
the limit $U \rightarrow \infty$
\cite{BHP}. A somewhat more sophisticated mean-field approximation, 
namely that based on the slave-boson formulation due to Kotliar and 
Ruckenstein \cite{KR}, is in principle not restricted to weak or 
strong coupling.
In fact, it was shown that this so-called slave-boson mean-field 
(SBMF) approximation is equivalent to the Gutzwiller approximation 
to the
Gutzwiller wave function \cite{Gutz}, the latter of which improves 
upon the HFA especially in the {\em intermediate}-coupling regime.
This SBMF approximation then allows to approach the interesting 
intermediate-coupling regime from the strong-coupling side.
Such an approximate strong-coupling approach can be particularly 
helpful since more rigorous Quantum Monte Carlo calculations
become increasingly cumbersome for stronger coupling \cite{QMC}.

In a previous work \cite{PDMB}, we compared the HFA and SBMF 
approximation for various simple magnetic phases, as well as 
computed the
effective hopping and spin stiffness for the {\em two-dimensional}
(one-band) Hubbard model. 
In this paper, I derive a large-$U$ expansion  
within the SBMF approximation for the Hubbard model 
in {\em arbitrary} dimension. Here, I restrict myself to the
antiferromagnetic phase at half-filling, but the extension
to doping the antiferromagnet with electrons or holes is 
straightforward (though tedious).

In Section 2, the slave-boson mean-field approximation is
briefly introduced and the resulting set of integral equations
for the antiferromagnetic phase is given. A large-$U$ expansion is
derived in Section 3 in terms of moments of the density of states
for freely hopping electrons on a lattice, and compared to 
earlier numerical and analytical work in the literature. In
Section 4, the expansion is made explicit for hypercubic
lattices of arbitrary dimension. Finally, I discuss the convergence
of the series and possible extensions. The Appendix contains
useful formulae for the moments of the above mentioned
density of states.

% Section 2: Preliminaries
\section{Slave-Boson Mean-Field Approximation}

The Hubbard Hamiltonian is \cite{Hub}: \\
\begin{equation}
{\cal H}_h = - t \sum_{j,\delta,\sigma } c_{j+\delta,\sigma}^{\dagger} 
c_{j \sigma}
+ U \sum_{j}n_{j \uparrow}n_{j \downarrow} ,   \label{eq:Hh}
\end{equation}
where $c_{j \sigma}$ is the 
annihilation operator for an electron at
site $j$ with spin $\sigma$. Neighboring 
sites of site $j$ are denoted by $j+\delta$. 
$t$ is the hopping integral,
$U$ the on-site interaction between electrons of opposite spin, and 
$n_{j \sigma}=c_{j \sigma}^{\dagger}c_{j \sigma}$ is the occupation 
number operator. The slave-boson approach of Kotliar and Ruckenstein
consists of introducing bosons for each of the (four) possible
electron occupancies of a site. The electron creation and 
annihilation operators are then modified such that always 
the (one) boson
corresponding to the electron occupancy is present at each site.
In a functional-integral description of the new Hamiltonian the
slave-boson mean-field (SBMF)
approximation is when the Bose fields are independent of (imaginary-) 
time. Different phases can be considered by assuming different
forms of position dependence of the Bose fields.
For more details see Refs.\cite{KR,PDMB}.
The SBMF approximation is an improvement over the
Hartree-Fock approximation since it takes some local correlations 
into account; in particular, the density of doubly occupied sites 
is an independent parameter to be optimized.

The set of equations describing the antiferromagnetic (AF) phase
within SBMF is given by \cite{PDMB}:
\begin{eqnarray}
m_{\rm s} & = & 2 \bar{\lambda} \int_{\bar{\mu}}^{\infty} 
\mbox{d}\varepsilon \,
\frac{{\cal N}(\varepsilon)}{\left(\varepsilon^2 + \bar{\lambda}^2 \right)^{1/2}}    \label{eq:ms} \\
\bar{\lambda} & = & \frac{2}{q_{\rm s}} [q_{\rm s}]_{m_{\rm s}} \int_{\bar{\mu}}^{\infty} 
\mbox{d}\varepsilon \, 
\frac{{\cal N}(\varepsilon) \varepsilon^2}{\left(\varepsilon^2 + 
\bar{\lambda}^2 \right)^{1/2}}    \label{eq:lab} \\
U & = & \frac{[q_{\rm s}]_d}{d} \int_{\bar{\mu}}^{\infty} 
\mbox{d}\varepsilon \,
\frac{{\cal N}(\varepsilon) \varepsilon^2}{\left(\varepsilon^2 + 
\bar{\lambda}^2 \right)^{1/2}}  
\label{eq:UU} \\
n & = & 2 \int_{\bar{\mu}}^{\infty} \mbox{d}\varepsilon \, 
{\cal N}(\varepsilon) ~,        \label{eq:nn} 
\end{eqnarray}
where $\bar{\lambda}$ is an internal (``renormalized gap'') parameter;
$\bar{\mu}$ an effective chemical potential; and 
the band renormalization $q_{\rm s}$ is a function of density 
$n$, sublattice magnetization $m_{\rm s}$, and $d$ 
($d^2$ is the density of doubly occupied sites), and
given by:
\begin{equation}
q_{\rm s} (n,m_{\rm s},d) = z(n,m_{\rm s},d) z(n,-m_{\rm s},d) ~, 
\label{eq:qs}
\end{equation}
where
\begin{equation}
z(n,m_{\rm s},d) =
\frac{\sqrt{ (1-n+d^2)(n + m_{\rm s} - 2d^2)} +  
d \sqrt{n - m_{\rm s} - 2d^2}}
 {\sqrt{(n + m_{\rm s}) \left( 1 - \frac{n + m_{\rm s}}{2} \right)} } ~. 
\label{eq:zsig}
\end{equation} 
$[q_{\rm s}]_{\alpha}$ is the derivative of $q_{\rm s}$ with respect 
to $\alpha$
(explicit expressions can be found in Ref.\cite{PDMB}) and
${\cal N}(\varepsilon)$ is the density of states (DOS) of 
freely hopping electrons.
The dimension-dependence resides solely in ${\cal N}(\varepsilon)$ 
and since this DOS
applies to a non-interacting system, I will be able to derive results
for arbitrary dimension in Section 4 (see also the appendix).
Note that the above set of equations reduces to the familiar
antiferromagnetic or spin-density wave (SDW) solution in the HFA
if $q_{\rm s} = 1$ \cite{SWZ}. 
In that case, one has one self-consistency equation (\ref{eq:ms}), 
which is the gap equation with
$\bar{\lambda}$ as gap parameter $\Delta$ ($\equiv Um_{\rm s}$),
instead of three self-consistency equations 
(\ref{eq:ms})--(\ref{eq:UU}).

The energy of the AF state (per site) 
and the spin stiffness $\rho_{\rm s}$ are given by:
\begin{eqnarray}
e_{\rm AF} & = & -2 q_{\rm s} \int_{\bar{\mu}}^{\infty} 
\mbox{d}\varepsilon 
\, {\cal N}(\varepsilon) \sqrt{\varepsilon^2 + 
\bar{\lambda}^2} + Ud^2 + \bar{\lambda} q_{\rm s} m_{\rm s}  
\label{eq:eaf} \\
\rho_{\rm s} & = & \frac{q_{\rm s}}{4} \int_{\bar{\mu}}^{\infty} \mbox{d}\varepsilon \,
\frac{{\cal N}(\varepsilon) \varepsilon^2}{\left(\varepsilon^2 + 
\bar{\lambda}^2 \right)^{1/2}} - 
\frac{z_+^2z_-^2}{q_{\rm s}}
\int_{\bar{\mu}}^{\infty} \mbox{d}\varepsilon \, 
\frac{{\cal N}_{\rm v}(\varepsilon) \varepsilon^2}{\left(\varepsilon^2 +
 \bar{\lambda}^2 \right)^{3/2}}          \label{eq:rhose}
\end{eqnarray}
${\cal N}_{\rm v}(\varepsilon)$ is the weighted DOS, which, 
as well as the combination $z_+z_-$, 
is specified in Ref.\cite{PDMB}. I will not repeat the expressions 
here 
since I will now restrict to half-filling ($\bar{\mu}=0$, $n=1$) 
for which case $z_+z_-$ equals zero. 
For half-filling, the band-renormalization reduces to:
\begin{equation}
 q_{\rm s}(1,m_{\rm s},d) = 
  \frac{2d^2 \left[ \sqrt{1-2d^2 + m_{\rm s}} + \sqrt{1-2d^2 - 
m_{\rm s}} \right]^2}
       {1 - m_{\rm s}^2} ~. \label{eq:qshf}
\end{equation}

% Section 3
\section{Large-$U$ expansion in terms of moments}

The above set of equations (\ref{eq:ms})--(\ref{eq:nn}) 
can be solved numerically if the DOS for freely
hopping electrons is given. For instance, in two dimensions ${\cal N}(\varepsilon)$ is
known analytically and numerical solutions to 
(\ref{eq:ms})--(\ref{eq:nn})
were obtained in Ref.\cite{PDMB}. Now I will derive solutions to
(\ref{eq:ms})--(\ref{eq:nn}) for $n=1$ in the form of a systematic
series expansion in $1/U$. This expansion is derived in two steps:
(i) From the explicit form of the equations above (or from numerical
solutions) it can be seen that large $U$ corresponds to large 
$\bar{\lambda}$.
Therefore, I first make a large-$\bar{\lambda}$ expansion for 
all quantities of interest. 
Subsequently, (ii) the $\bar{\lambda}$-expansion for $U$ 
(following from equation (\ref{eq:UU})) is inverted and substituted
in the large-$\bar{\lambda}$ expansion for all the relevant quantities 
obtained before. 
In this way, one obtains the desired large-$U$ expansion.

By expanding the square root in the denominator of the integrals
in (\ref{eq:ms})--(\ref{eq:nn}) for 
$\bar{\lambda} \gg \varepsilon$ one obtains
a large-$\bar{\lambda}$ expansion in terms of {\em moments}
$M_n$ of the DOS:
\begin{equation}
 M_n = \int_{-\infty}^{\infty} \mbox{d}\varepsilon \, 
{\cal N}(\varepsilon) \varepsilon^n ~. \label{eq:Mn}
\end{equation}
After some work one arrives at the following 
expansions in $q \equiv 1/\bar{\lambda}$ \cite{Exp}:
\begin{eqnarray}
 m_{\rm s} & = & 1 - {\scriptstyle \frac{1}{2}} M_2 \, q^2 + 
{\scriptstyle \frac{3}{8}} M_4 \, q^4 - 
{\scriptstyle \frac{5}{16}} M_6 \, q^6 + 
{\scriptstyle \frac{35}{128}} M_8 \, q^8 
+ {\cal O}(q^{10})  ~, \\
   d & = & {\scriptstyle \frac{1}{2}} \sqrt{M_2} \, q - 
\frac{4 M_2^2 + 3 M_4}{16\sqrt{M_2}} \, q^3 + 
\frac{\sqrt{M_2}}{16} \left[
M_2^2 + {\scriptstyle \frac{13}{2}} M_4 -  \frac{9M_4^2}{16M_2^2} + 
 \frac{5M_6}{2M_2} \right] q^5 \nonumber \\
 ~~ & ~~ &  \frac{-1}{2048M_2^{5/2}} \left[ 192 M_2^6
+ 368 M_2^4 M_4 + 364 M_2^2 M_4^2 + 27 M_4^3 + \right. \nonumber \\
 ~~ & ~~ & \left. 736 M_2^3 M_6 - 
120 M_2 M_4 M_6 + 280 M_2^2 M_8 \right] q^7 + {\cal O}(q^{9}) ~,
 \label{eq:xd} \\
 q_{\rm s} & = & 1 - \frac{M_2}{4} \, q^2 - 
\left[ \frac{ M_2^2 - 7 M_4}{16} \right] q^4  \nonumber \\
~~ & ~~ & - \left[ \frac{9 M_2^3 - 10 M_2 M_4 + 12 M_4^2/M_2 + 
26 M_6}{64} \right] q^6 + {\cal O}(q^{8}) ~, \\
   U & = & \frac{2}{q} + {\scriptstyle \frac{5}{2}} M_2 \, q + \left[ 
\frac{11 M_2^2 - 21 M_4}{8} \right] q^3 + {\cal O}(q^{5}) ~, \\
 e_{\rm AF} & = & - {\scriptstyle \frac{1}{2}} M_2 \, q + 
\left[ \frac{3 M_2^2 + M_4}{8} 
\right] q^3 + \left[ \frac{M_2^3 - 22 M_2 M_4 - 2 M_6}{32}
\right] q^5 + {\cal O}(q^{7}) ~, \\
 \rho_{\rm s} & = & \frac{M_2}{8} \, q - 
\left[ \frac{M_2^2 + 2 M_4}{32}  \right] q^3 
 + \left[ \frac{- M_2^3 + 9 M_2 M_4 + 6 M_6}{128} \right] q^5  
\nonumber \\
~~ & ~~ & - \left[ \frac{9 M_2^4 - 12 M_2^2 M_4 + 26 M_4^2 + 
32 M_2 M_6 +20 M_8}{512} \right] q^7 + {\cal O}(q^{9})  ~.
\end{eqnarray}
For notational convenience the hopping integral $t$ is
taken equal to 1 in this section.
For $m_{\rm s}$, $d$, and $q_{\rm s}$ I have listed precisely 
the number of 
terms that is needed in order to obtain the number of terms 
(three) that is
listed for $U$. For $e_{\rm AF}$ and $\rho_{\rm s}$ the terms are 
given that can be 
found using the listed terms for $m_{\rm s}$, $d$, 
$q_{\rm s}$, and $U$.
Note that this implies that, to obtain the leading order in 
$q$ for $U$,
one already needs three terms for $m_{\rm s}$ and two terms for $d$.

Inverting the expansion for $U$ gives:
\begin{equation}
 q = 2p + 10 M_2 \, p^3 + (122 M_2^2 - 42 M_4) p^5 + 
{\cal O}(p^{7}) ~,
\end{equation}
where I have introduced the notation: $p \equiv 1/U$. 
Substituting this
into the $q$-expansions, one obtains the following 
large-$U$ expansions:
\begin{eqnarray}
 m_{\rm s} & = & 1 - 2 M_2 \, p^2 - \left[ 20 M_2^2 - 6 M_4 \right] 
p^4 \nonumber \\
 ~~ & ~~ & - \left[ 294 M_2^3 - 204 M_2 M_4 + 20 M_6 \right] p^6 + 
{\cal O}(p^{8}) ~, \\
 d^2 & = & M_2 \, p^2 + \left[ 6 M_2^2 - 3 M_4 \right] p^4 +
  \left[ 75 M_2^3 - 70 M_2 M_4 + 10 M_6 \right] p^6 + 
{\cal O}(p^{8}) ~, \\
 q_{\rm s} & = & 1 - M_2 \, p^2 - \left[ 11 M_2^2 - 7 M_4 \right] 
p^4  \nonumber \\
~~ & ~~ & - \left[ 176 M_2^3 - 192 M_2 M_4 + 12 M_4^2/M_2 + 
26 M_6 \right] p^6 + {\cal O}(p^{8}) ~, \\
 e_{\rm AF} & = & - M_2 \, p - \left[ 2 M_2^2 - M_4 \right] p^3 -
\left[ 15 M_2^3 - 14 M_2 M_4 + 2 M_6 \right] p^5 + {\cal O}(p^{7}) 
\label{eq:eafx} ~, \\
 \rho_{\rm s} & = & \frac{M_2}{4} \, p + 
{\scriptstyle \frac{1}{2}} \left[2 M_2^2 - M_4 \right] p^3 +
{\scriptstyle \frac{3}{4}} \left[ 15 M_2^3 - 14 M_2 M_4 + 
2 M_6 \right] p^5 + {\cal O}(p^{7}).
\end{eqnarray}
A consistency check on the above results is that the following 
general identity still holds for the expansions above:
\begin{equation}
 \frac{\partial e_{\rm AF}}{\partial U} = d^2 ~. \label{eq:d2id}
\end{equation}
Because in the SBMF at half-filling the spin stiffness is given by
$- {\scriptstyle \frac{1}{8}}$ times the average kinetic energy
\cite{PDMB}, we also have:
\begin{equation}
 \rho_{\rm s} = - {\scriptstyle \frac{1}{8}} \left( e_{\rm AF} - 
U d^2 \right) ~.
\end{equation}
Combining the latter formula with (\ref{eq:d2id}) it immediately
follows that the expansion for $\rho_{\rm s}$ follows directly 
from the expansion for $e_{\rm AF}$, as is also seen in the way 
the expansions are written above.

It is of some interest to compare the present result with results
from the HFA and other expressions in the literature. It is far
easier to derive the corresponding expression to (\ref{eq:eafx}) 
in the HFA since there is only one consistency equation (the gap 
equation) instead of three, and there is no band-renormalization
function $q_{\rm s}$ which requires perturbative expanding.
Following the same procedures as for SBMF, I find for the HFA:
\begin{equation}
 e_{\rm AF} = - M_2 \, p - \left[ M_2^2 - M_4 \right] p^3 -
 \left[ 4 M_2^3 - 6 M_2 M_4 + 2 M_6 \right] p^5 + {\cal O}(p^{7}) ~.
\label{eq:eafHF}
\end{equation}
Comparing with (\ref{eq:eafx}), one sees that HFA and SBMF 
give the same
coefficient for the leading order contribution, but that there are 
differences for the higher order contributions. Note that also 
for the
higher order contributions the term in the coefficient involving the 
highest moment agrees between HFA and SBMF. Since the moments are
always positive, it is also clear that for large $U$ the energy 
in the SBMF approximation is always lower than in the HFA.

In dimensions $D=2$ and $D=3$ (square and simple cubic lattices),
the results can also be compared with a large-$U$ expansion by
Takahashi \cite{Taka}. He derives a rigorous expansion
(i.e. without making approximations) for the half-filled Hubbard
model in terms of spin correlation
functions of the $s = {\scriptstyle \frac{1}{2}}$ Heisenberg model. 
These correlation functions
are then evaluated using the Linear Spin Wave (LSW) approximation.
Below the results of the latter approach are compared to those of
HFA and SBMF for $D=2$ and $D=3$ (the necessary moments are easily
evaluated and may for instance be found in the appendix):
\begin{eqnarray}
      D=2 ~~~~~~~~~~~~~ & ~~~ &~~~~~~~~~~ D=3    \nonumber \\
\mbox{HFA}~~~~~ -4 p + 20 p^3 - 192 p^5 + \cdots  & ~~~ & 
          -6 p + 54 p^3 - 1344 p^5 + \cdots \nonumber \\
\mbox{SBMF}~~~~ -4 p + ~4 p^3 + 256 p^5 + \cdots  & ~~~ & 
          -6 p + 18 p^3 + ~600 p^5 + \cdots \label{eq:Cf} \\
\mbox{LSW}~~~~~ -4.63 p + 34.6 p^3  + \cdots ~~~~~ & ~~~ & 
          -6.58 p + 65.6 p^3  + \cdots \nonumber
\end{eqnarray}
It is clear that for large enough $U$ the LSW expansion will 
give the lower energy. 
The leading order coefficient is proportional to the
ground-state energy of the $s = {\scriptstyle \frac{1}{2}}$ 
Heisenberg antiferromagnet, which
is known to be very well approximated by LSW \cite{And,Mattis}.
Both the HFA and SBMF do not do better than the mean-field result 
for the leading order coefficient.
In view of the fact that SBMF is an improvement over the HFA, it is
somewhat remarkable that the HFA results for the next-to-leading order
resemble the LSW results more than the SBMF results do. 
Also the different sign of the coefficient of $p^5$ between 
HFA and SBMF
is notable. Further discussion of the range of validity 
of the expansion is given below.

To end this section a quantitative discussion of the large-$U$
expansion is also given. In Table I, the ground-state energy 
of the AF phase
at half-filling as obtained from the large-$U$ expansion within SBMF
($e_{\rm AF}^{\rm Uexp}$) is compared to the 
``exact \mbox{SBMF}'' result
[$e_{\rm AF}^{\rm SBMF}$, found by solving (\ref{eq:ms})--(\ref{eq:nn})
numerically] and to variational Monte Carlo results using
an (antiferromagnetic) Gutzwiller wave function 
($e_{\rm AF}^{\rm GWVMC}$,
from Ref.\cite{YS}) for $D=2$. For $D=3$, only 
$e_{\rm AF}^{\rm Uexp}$ and 
$e_{\rm AF}^{\rm GWVMC}$ are compared.
Note that the large-$U$ expansion approximates the 
``exact SBMF'' result
very well for $U/t \geq 7$; it is expected that 
similar agreement is found
for $D=3$, where it is more involved to compute 
$e_{\rm AF}^{\rm SBMF}$
since ${\cal N}(\varepsilon)$ is not a known analytical function 
as it is for $D=2$ \cite{Has}.
This is consistent with the fact that
from the coefficients in the explicit expansions for $D=2$ and $D=3$,
formula (\ref{eq:Cf}), one would estimate the series to converge 
well for $U/t \geq 8$ and $U/t \geq 6$ for two and three 
dimensions, respectively.
SBMF is equivalent to the Gutzwiller {\em approximation} to the
Gutzwiller wave function \cite{KR}
and this approximation becomes exact in the
limit of an infinite number of spatial dimensions \cite{MV,Geb}. 
From the
comparison with the GWVMC results it can be seen that 
the approximation
is quite good already for $D=2$. In view of the facts, that 
for $D=3$ a very
small lattice was used in the GWVMC calculations and a 
larger lattice will
even raise the energy somewhat (the effect will be 
larger for smaller $U$),
the agreement for $D=3$ can even be 
called excellent. So the clear advantage of the large-$U$ expansion
is that it can give results that are, for $U/t \geq 8$,
within $1 \%$ of the ``exact
SBMF'' and GWVMC calculations (which especially for $D=3$ are much
more involved and computationally demanding) by means of 
the simple formula (\ref{eq:eafx}).
\\[0.2cm]
\begin{center}
\begin{tabular}{|r|rcc|cc|} \hline \hline
\multicolumn{1}{|c}{} & \multicolumn{3}{|c}{} & 
\multicolumn{2}{|c|}{} \\
\multicolumn{1}{|c}{} & \multicolumn{3}{|c}{$D=2$} & 
\multicolumn{2}{|c|}{$D=3$} \\
\multicolumn{1}{|c}{} & \multicolumn{3}{|c}{} & 
\multicolumn{2}{|c|}{} \\
~$U$~~ & ~$-e_{\rm AF}^{\rm SBMF}$ & $-e_{\rm AF}^{\rm Uexp}$ & 
$-e_{\rm AF}^{\rm GWVMC}$ & ~~$-e_{\rm AF}^{\rm Uexp}$ & 
$-e_{\rm AF}^{\rm GWVMC}$ \\ 
\multicolumn{1}{|c}{} & \multicolumn{3}{|c}{} & 
\multicolumn{2}{|c|}{} \\
\hline
\multicolumn{1}{|c}{} & \multicolumn{3}{|c}{} & 
\multicolumn{2}{|c|}{} \\
 ~ 6~~  &  ~~0.623639  &  0.615226  &  0.629(3)  &  ~~~0.8395 &
  0.886(5) \\
 ~ 8~~  &  ~~0.485104  &  0.484375  &  0.493(3)  &  ~~~0.6965 & 
 0.704(7) \\
 ~10~~  &  ~~0.393528  &  0.393440  &  0.401(4)  &  ~~~0.5760 & 
 0.579(6)  \\
 ~12~~  &  ~~0.330002  &  0.329990  &  0.336(5)  &  ~~~0.4872 &
  0.491(7)  \\
 ~16~~  &  ~~0.248782  &  0.248779  &    -       &  ~~~0.3700 & 
   -      \\
 ~20~~  &  ~~0.199420  &  0.199420  &    -       &  ~~~0.2976 & 
   -      \\
\multicolumn{1}{|c}{} & \multicolumn{3}{|c}{} & 
\multicolumn{2}{|c|}{} \\
 \hline \hline 
\end{tabular}\\[0.4cm]
\end{center}
Table I. Numerical results in $D=2$ and $D=3$ for the ground-state
energy of the antiferromagnetic phase at half-filling. 
$e_{\rm AF}^{\rm Uexp}$ is evaluated using the large-$U$ expansion
(\ref{eq:eafx}) and $e_{\rm AF}^{\rm SBMF}$ is obtained from
(\ref{eq:eaf}) after solving (\ref{eq:ms})--(\ref{eq:nn}).
The GWVMC results are taken from Ref.\cite{YS} and were obtained
on $20\times 20$ and $6\times 6\times 6$ lattices for 
$D=2$ and $D=3$, respectively. 
Between brackets the statistical error in the
last digit is given.

% Section 4
\section{Large-$U$ expansion for arbitrary dimension}

In the appendix, it is shown that the moments $M_n$ can be obtained
as a function of dimension $D$ [formula (\ref{eq:MnD})]. 
If one substitutes
these expressions in (\ref{eq:eafx}) and (\ref{eq:eafHF})
one obtains a simultaneous $1/U$ and
$1/D$ expansion for the energy of the AF phase:
\begin{eqnarray}
\mbox{SBMF}~~~ e_{\rm AF} & = & - \frac{2Dt^2}{U} + 
\left( 1 - \frac{3}{2D} \right)
\frac{4D^2 t^4}{U^3} + \left( -~3 + \frac{24}{D} - 
\frac{20}{D^2} \right)
\frac{8D^3 t^6}{U^5} + \cdots ~,  \\
\mbox{HFA}~~~~ e_{\rm AF} & = & - \frac{2Dt^2}{U} + 
\left( 2 - \frac{3}{2D} \right)
\frac{4D^2 t^4}{U^3} + \left( -16 + \frac{36}{D} - 
\frac{20}{D^2} \right)
\frac{8D^3 t^6}{U^5} + \cdots ~, 
\end{eqnarray}
where I have included the hopping parameter $t$ again. 
The above formulas
are well suited to discuss the limit of an infinite number of
dimensions. This is of interest since, as noted above,
SBMF is equivalent to the Gutzwiller approximation to the
Gutzwiller wave function; an approximation which becomes 
exact in the
limit of an infinite number of spatial dimensions. 
Thus the  present expansion
allows us to scrutinize the approximation explicitly 
for lower dimension. 
In order for the
limit $D \rightarrow \infty$ to be meaningful, one has to 
introduce a
scaled hopping parameter $t^{\ast} = t \sqrt{2D}$ \cite{MV}. 
In terms of
this scaled parameter the above SBMF large-$U$ expansion 
assumes the form:
\begin{equation}
\frac{e_{\rm AF}}{t^{\ast}} = - \frac{t^{\ast}}{U} + \left( 1 - 
\frac{3}{2D} \right)
\left(\frac{t^{\ast}}{U} \right)^3 + \left( - 3 + 
\frac{24}{D} - \frac{20}{D^2} \right) \left(\frac{t^{\ast}}{U} 
\right)^5 + \cdots   \label{eq:arbD}
\end{equation}
In Table II, I evaluate (\ref{eq:arbD}) for $D=2, 3, 4$, and 
$\infty$.
\\[0.5cm]
\begin{center}
\begin{tabular}{|r|rccc|} \hline \hline
\multicolumn{1}{|c}{} & \multicolumn{4}{|c|}{} \\
~$U^{\ast}$~~ & $-e_{\rm AF}^{\ast}(D=2)$ & 
             $-e_{\rm AF}^{\ast}(D=3)$ & 
             $-e_{\rm AF}^{\ast}(D=4)$ &
             $-e_{\rm AF}^{\ast}(D=\infty)$  \\
\multicolumn{1}{|c}{} & \multicolumn{4}{|c|}{} \\
\hline
\multicolumn{1}{|c}{} & \multicolumn{4}{|c|}{} \\
 ~ 2~~  &  0.34375  &  0.35069  &  0.36719  &  0.46875  \\
 ~ 4~~  &  0.24219  &  0.23947  &  0.23853  &  0.23730  \\
 ~ 8~~  &  0.12439  &  0.12394  &  0.12373  &  0.12314  \\
 ~12~~  &  0.08317  &  0.08303  &  0.08296  &  0.08277  \\
 ~16~~  &  0.06244  &  0.06238  &  0.06235  &  0.06226  \\
 ~20~~  &  0.04997  &  0.04994  &  0.04992  &  0.04988  \\
\multicolumn{1}{|c}{} & \multicolumn{4}{|c|}{} \\ \hline \hline
\end{tabular}\\[0.5cm]
\end{center}
Table II. Scaled ground-state energy of the AF phase at half-filling
$e_{\rm AF}^{\ast} \equiv e_{\rm AF}/t^{\ast}$
within SBMF as a function of scaled interaction 
$U^{\ast} \equiv U/t^{\ast}$
for various dimensions $D$ (with $t^{\ast} = t \sqrt{2D}$).
\\[0.5cm]

From Table II, one sees that for scaled interaction 
$U^{\ast} \geq 4$
(which corresponds to $U/t = 8$ for $D=2$) already the result for
$D=2$ and the infinite-dimensional result coincide remarkably 
well. The
simultaneous large-$U$, large-$D$ expansion shows 
in detail
how well the Gutzwiller approximation, and therefore SBMF, 
reproduces 
the Gutzwiller wave function for the various dimensions. 
It would have been
interesting to compare the coefficients in the present results
with those obtained by Gebhard in his $1/D$-expansion for the
half-filled Hubbard model \cite{Geb}.
Unfortunately, a direct comparison is not meaningful, since
Gebhard only considers the paramagnetic Gutzwiller wave function
and does not treat the antiferromagnetic phase at half-filling in
a $1/D$-expansion.

% Section 
\section{Discussion and Conclusions}

There is no fundamental reason why the large-$U$ expansion 
given here 
cannot be extended to higher orders in $t/U$. With a little more
effort (especially the coefficients in the large-$\bar{\lambda}$ 
expansion of $d$ in
formula (\ref{eq:xd}) require some care) more terms can be
obtained. However, as already remarked in Section 3,
the present series reproduces well the ``exact SBMF'' result for
$U/t \geq 7$ in two dimensions, and this agreement is expected to
be even better in three dimensions. Also, for smaller values 
of $U/t$
the contribution from the $(t/U)^5$ term becomes larger than
that from the $(t/U)^3$ term, which is an indication that the
intrinsic radius of convergence of the series is close to
$U/t = 7$. Therefore, extending the series can not be considered 
very useful.

A more interesting extension of the present series is to go off
half-filling, i.e. $n \neq 1$. Again there is no fundamental
reason preventing this, but now the task is quite a bit more
laborious for a number of reasons. Not only are we dealing with
the case $\bar{\mu} \neq 0$ and consequently {\em partial} moments
are required [see (\ref{eq:Mn})], which can only be computed
numerically, also the expression for the band renormalization
$q_{\rm s}$, formulae (\ref{eq:qs})--(\ref{eq:zsig}), needs to 
be expanded for
$n \neq 1$, which implies that non-rational coefficients in
the expansion in terms of (partial) moments appear.
Furthermore, in formula (\ref{eq:rhose}) for $\rho_{\rm s}$ also the 
term with the combination $z_+z_-$ needs to be taken into 
account and perturbatively expanded. 
If similar large-$U$ expansions are
possible for more complicated phases, like spirals or domain
walls, this would allow for a study of aspects of the phase 
diagram off half-filling.

In conclusion, a large-$U$ expansion is derived for the
antiferromagnetic phase of the Hubbard model within the
framework of the slave-boson mean-field approximation.
Even though such an expansion constitutes in a sense an
approximation of an approximation, the resulting analytic
expression is capable of
reproducing very well results of elaborate Monte Carlo
calculations. Furthermore, the expansion allows for
an explicit study of the limit of a large number of
spatial dimensions, since the occurring moments of the
non-interacting density of states are obtained for
arbitrary dimension.

\section*{Acknowledgments}
I acknowledge useful discussions with P.G.J. van Dongen,
F. Gebhard, and
J.M.J. van Leeuwen on the work reported here, as well as
comments by D.P. Aalberts on an earlier version of the paper.

%Appendix
\newpage
\appendix
\section*{Appendix}
\renewcommand{\theequation}{A.\arabic{equation}}
\setcounter{equation}{0}
In this appendix some general expressions are derived for the 
moments
$M_n$ of the density of states ${\cal N}(\varepsilon)$ of 
freely hopping electrons on hypercubic lattices:
\begin{equation}
 M_n = \int_{-\infty}^{\infty} \mbox{d}\varepsilon \, 
{\cal N}(\varepsilon) \varepsilon^n 
     = \frac{(2t)^n}{(2\pi)^D} \int_{-\pi}^{\pi} \cdots 
\int_{-\pi}^{\pi}
        \mbox{d}k_1 \ldots \mbox{d}k_D \, 
        \left[ \cos k_1 + \cdots + \cos k_D \right]^n  ~. 
\label{eq:MnA}
\end{equation}
Although such moments are frequently used
and calculated in the literature I have never encountered the
general analytic expressions and simple formulae to be given below.
First, I will derive simple expressions for {\em all} moments
in the {\em lower} dimensions $D=1, 2$, and $3$. Then I will derive 
expressions for the {\em lower} moments in {\em arbitrary} 
dimension $D$.

By employing the multinomial generalization of the binomial it is
possible to further work out (\ref{eq:MnA}):
\begin{equation}
 M_n = (2t)^n \sum_{n_1,\ldots,n_D = 0}^{n} \hspace{-0.5cm} \raisebox{.8ex}{$\prime$} 
       \hspace{0.5cm} \frac{n!}{n_1!\ldots n_D!}
        q_{n_1}\ldots q_{n_D} ~, \label{eq:MnA2}
\end{equation}
where the prime on the sum denotes the restriction: 
$n_1 + \cdots + n_D = n$ and
\begin{equation}
 q_m = \frac{1}{2\pi} \int_{0}^{2\pi} \mbox{d}k \, 
\cos^m (k) = \left\{
       \begin{array}{cl} \frac{(m-1)!!}{m!!} & ~~\mbox{$m$ even} \\
                               0             & ~~\mbox{$m$ odd} 
\end{array}
       \right. 
\end{equation}
From this general form it is clear that all moments with 
$n$ odd are zero
for any dimension. All formulae below will be for moments 
with $n$ even.
For $D=1$ the restriction allows for only one term in the 
sum and the result is trivial:
\begin{equation}
   M_n = (2t)^n \frac{(n-1)!!}{n!!}  ~~~~~~D=1 ~.
\end{equation}
A more remarkable result is that the moments for $D=2$ are 
exactly the
squares of the moments for $D=1$ (if $t$ is put equal to 1):
\begin{equation}
   M_n = t^n \left[ 2^n \frac{(n-1)!!}{n!!} \right]^2 ~~~~~~D=2 ~.
 \label{eq:Mn2D}
\end{equation}
The proof of (\ref{eq:Mn2D}) follows directly from (\ref{eq:MnA2}) 
by some manipulations with the factorials and employing the
identity \cite{GR}:
\begin{equation}
  \sum_{n=0}^{N} {N \choose n}^2 = {2N \choose N}  ~. 
\label{eq:GRid}
\end{equation}
For $D=3$ I have not been able to obtain an explicit expression for
$M_n$, but the triple sum in (\ref{eq:MnA2}) can be reduced to
one simple, unrestricted sum. Using the relations: 
$(m-1)!!/m! = 1/m!! =
2^{-m/2}/(m/2)!$ ($m$ even), one can rewrite (\ref{eq:MnA2}) 
(for $n$ even) as:
\begin{equation}
 M_n = n! \sum_{u,v,w = 0}^{n/2} \hspace{-0.3cm} 
\raisebox{.8ex}{$\prime$} \hspace{0.4cm}
       \frac{1}{(u!)^2(v!)^2(w!)^2}  ~, \label{eq:BR}
\end{equation}
where the prime indicates the constraint: $u + v + w = n/2$. 
This expression
was already given by Brinkman and Rice for the moments of the
density of states of one hole in a ferromagnetic background 
\cite{BR}.
It can however be simply reduced to a single sum by 
incorporating the
constraint and a subsequent application of (\ref{eq:GRid}). 
The result is:
\begin{equation}
 M_n = n! \, \sum_{u=0}^{n/2} \, 
          \frac{(n-2u)!}{(u!)^2 \left( (n/2 - u)! \right)^4} 
~~~~~~D=3 ~. \label{eq:Mn3D}
\end{equation}
Formulas (\ref{eq:Mn2D}) and (\ref{eq:Mn3D}) make for much easier 
evaluation of the moments than is usually found in the literature:
e.g. in Ref.\cite{Haaf} moments up to $n=22$ and $n=16$ are computed
for $D=2$ and $D=3$, respectively, using the method of counting
the number of paths that return to the origin after $n$ steps.
I note that also finding an explicit expression for $D=3$ may be
possible on account of the (empirical) fact that the moments 
for $D=3$
always contain the moments for $D=1$ as a factor.

The general expression (\ref{eq:MnA2}) shows that finding the 
terms contributing to $M_n$ amounts to finding all the combinations 
of $D$ even numbers (including zero) that add up to $n$. 
For the lower moments the number
of possibilities is not large and can be expressed in terms of $D$
using combinatorial arguments. I have found the following results:
\begin{eqnarray}
 M_2 & = & 2D  \nonumber \\
 M_4 & = & 6D(2D-1) \nonumber \\
 M_6 & = & 20D(6D^2 - 9D + 4) \label{eq:MnD} \\
 M_8 & = & 70D(24D^3 - 72D^2 + 82D - 33) \nonumber \\
 M_{10} & = & 252D(14400D^4 - 143400D^3 + 501050D^2 - 
715225D + 343176) 
\nonumber
\end{eqnarray}


\begin{thebibliography}{99}
\bibitem{QMC} The sign problem is more severe for large $U/t$ 
(off half-filling) and most Quantum Monte Carlo work is for
two dimensions (where $W = 8t$) and $U/t = 4$.
\bibitem{Exp} The expansion for $m_{\rm s}$ follows directly 
from (\ref{eq:ms}),
the expansion for $d$ is first written with general coefficients 
which are then determined by the fact that (\ref{eq:lab}) must be 
satisfied
(this requires an expansion of $[q_{\rm s}]_{m_{\rm s}}$). 
The expansions
for $U$, $e_{\rm AF}$ and $\rho_{\rm s}$ then follow from those for 
$m_{\rm s}$ and $d$
(they require expansions of $[q_{\rm s}]_d$ and $q_{\rm s}$).
\end{thebibliography}
\end{document}